\documentclass[ALICE,manyauthors]{cernphprep}
\usepackage{ALICE_CDS_cernphprep}
\DeclareUnicodeCharacter{00A0}{}
\begin{document}

\begin{titlepage}
\PHyear{2020}
\PHnumber{186}
\PHdate{1 October}

\title{Production of muons from heavy-flavour hadron decays at high transverse momentum in \PbPb{} collisions at $\sNN=5.02$ and $2.76$~\TeV}
\ShortTitle{Production of muons from heavy-flavour hadron decays in Pb--Pb collisions}

\Collaboration{ALICE Collaboration\thanks{See Appendix~\ref{app:collab} for the list of collaboration members}}
\ShortAuthor{ALICE Collaboration}

\begin{abstract}
Measurements of the production of muons from heavy-flavour hadron decays in \PbPb{} collisions at $\sNN$~$=$~$5.02$ and~$2.76$~\TeV{} using the ALICE detector at the LHC are reported.
The nuclear modification factor $\RAA$ at \fivenn{} is measured at forward rapidity ($2.5 < y <4$) as a function of transverse momentum $\pT$ in central, semi-central, and peripheral collisions over a wide $\pT$ interval, $3 < \pT < 20$~\GeVc{}, in which muons from beauty-hadron decays are expected to take over from charm as the dominant source at high $\pT$ ($\pT > 7$~GeV/$c$). 
%With a significantly improved precision compared to the measurements at lower collision energy, 
The $\RAA$ shows an increase of the suppression of the yields of muons from heavy-flavour hadron decays with increasing centrality.
A suppression by a factor of about three is observed in the $10\%$ most central collisions.
The $\RAA$ at \fivenn{} is similar to that at \twosevensixnn{}. The precise $\RAA$ measurements have the potential to distinguish between model predictions implementing different mechanisms of parton energy loss in the high-density medium formed in heavy-ion collisions. They place important constraints for the understanding of the heavy-quark interaction with the hot and dense QCD medium.
%The results place stringent constraints on the relative energy loss between charm and beauty quarks.

\end{abstract}
\end{titlepage}

\setcounter{page}{2}
\section{Introduction}\label{sec:intro}

The study of ultra-relativistic heavy-ion collisions aims to investigate a state of strongly-interacting matter at high energy density and temperature.
Under these extreme conditions, quantum chromodynamics (QCD) calculations on the lattice predict the formation of a quark--gluon plasma (QGP), where quarks and gluons are deconfined, and chiral symmetry is partially restored~\cite{Borsanyi:2010bp,Bazavov:2011nk,Borsanyi:2013bia,Bazavov:2018mes}.

Heavy quarks (charm and beauty) are key probes of the QGP properties in the laboratory.
They are predominantly created in hard-scattering processes at the early stage of the collision on a timescale shorter than the formation time of the QGP of $\sim 0.1$--$1$ fm/$c$~\cite{Andronic:2015wma,Liu:2012ax}.
Therefore, they experience the full evolution of the hot and dense QCD medium.
During their propagation through the medium, they lose energy via radiative and collisional processes~\cite{Baier:1996sk,Dokshitzer:2001zm,Djordjevic:2003zk,Zhang:2003wk,Wicks:2007am,Gossiaux:2010yx}.
Quarks are expected to lose less energy than gluons due to the colour-charge dependence of the strong interaction.
Furthermore, several mass-dependent effects can also influence the energy loss.
Due to the dead-cone effect~\cite{Dokshitzer:2001zm,Djordjevic:2003zk,Armesto:2003jh}, the heavy-quark radiative energy loss is reduced compared to that of light quarks and the energy loss of beauty quarks is expected to be smaller than that of charm quarks.
The collisional heavy-quark energy loss is also expected to be reduced since the spatial diffusion coefficient, which controls the momentum exchange with the medium, is predicted to scale with the inverse of the quark mass~\cite{vanHees:2005wb}.
In addition to the heavy-quark energy loss, modifications of the hadronisation process via fragmentation and/or recombination~\cite{Greco:2003vf,Andronic:2003zv} and initial-state effects such as the modification of the parton distribution functions (PDF) inside the nucleus~\cite{Eskola:2009uj,Kopeliovich:2002yh,Vitev:2007ve} can also change the particle yields and phase-space distributions.
The medium effects can be quantified using the nuclear modification factor $\RAA$, which is the ratio between the $\pT$- and $y$-differential particle yields in nucleus-nucleus (AA) collisions ($\dd^{2} N_{\rm AA}/\dd\pT\dd y$) and the corresponding production cross section in pp collisions ($\dd^{2}\sigma_{\rm pp}/\dd\pT\dd y$) scaled by the average nuclear overlap function $\avg{\TAA}$:
\begin{equation}
\RAA(\pT, y) =
\frac{1}{\avg{\TAA}} \times
\frac{\dd^{2} N_{\rm AA}/\dd\pT\dd y}{\dd^{2}\sigma_{\rm pp}/\dd\pT\dd y}.
\end{equation}
The $\avg{\TAA}$ is defined as the ratio between the average number of nucleon--nucleon collisions $\avg{\Ncoll}$ and the inelastic nucleon--nucleon cross section~\cite{ALICE-PUBLIC-2018-011}.

Evidence of a strong suppression of open heavy-flavour yields was observed in central Au--Au and Cu--Cu collisions at $\sNN$~$=$~$200$~\GeV by the PHENIX and STAR collaborations at RHIC and in Pb--Pb collisions at \twosevensixnn by the ALICE, ATLAS, and CMS collaborations at the LHC (see~\cite{Andronic:2015wma} and references therein, and~\cite{Aaboud:2018bdg,Acharya:2018upq, Adam:2018inb,Adam:2016khe}).
Recently, the ALICE and CMS collaborations reported a significant suppression of the prompt D-meson yields measured at midrapidity in the $10\%$ most central \PbPb{} collisions at \fivenn with respect to the scaled pp reference, reaching a factor of about $5$--$6$ in the interval $8 < \pT < 12$~\GeVc~\cite{Acharya:2018hre,Sirunyan:2017xss}.
A strong suppression of the yields of high-$\pT$ electrons from heavy-flavour hadron decays was also observed by the ALICE collaboration at midrapidity in the \cent{0}{10} centrality class, where the measured $\RAA$ is about $0.3$ at $\pT\sim 7$~\GeVc~\cite{Acharya:2019mom}.
The suppression is similar to that observed for prompt D mesons and leptons from heavy-flavour hadron decays at \twosevensixnn~\cite{Adam:2015sza,Adam:2016khe,Aaboud:2018bdg}.
The nuclear modification factor of $\Bpm$ mesons, reconstructed via the exclusive decay channel $\Bpm\to\Jpsi\Kpm\to\mup\mum\Kpm$ with the CMS detector for $\abs{y} < 2.4$ and $7 < \pT < 50$~\GeVc, indicates a suppression of about a factor two in \PbPb{} collisions (\cent{0}{100} centrality class) at \fivenn~\cite{Sirunyan:2017oug} compatible with that of J/$\psi$ from b-hadron decays (non-prompt J/$\psi$)~\cite{Sirunyan:2017isk}.
A similar suppression as for $\Bpm$ mesons and non-prompt \jpsi{} is also observed for non-prompt $\Dzero$ mesons in the kinematic region $\abs{y}< 2.4$ and $2 < \pT < 100$~\GeVc~\cite{Sirunyan:2018ktu}.
The suppression of B mesons is weaker than that of prompt $\Dzero$ mesons at about $\pT = 10$~\GeVc, in line with the expected quark-mass ordering of energy loss.

This letter presents the first measurement of open heavy-flavour production via muons from semi-leptonic decays of charm and beauty hadrons in \PbPb{} collisions at \fivenn with the ALICE detector at the LHC. These measurements are carried out in the forward rapidity region ($2.5 < y < 4$), presently only covered by the ALICE experiment at the LHC in Pb--Pb collisions. They extend the measurement of open heavy-flavour production from mid to forward rapidities, providing a tomography of the QGP medium in broader phase space region. The analysis of muon-triggered events and large branching ratios ($\sim 10\%$) allow us to perform high precision measurements of the $\pT$-differential $\RAA$ of these muons over a broad $\pT$ interval, extended for the first time to $\pT = 20$~\GeVc in central ($0-10\%$), semi-central ($20-40\%$), and peripheral ($60-80\%$) collisions. This gives access to the investigation of medium effects in a new kinematic regime where the contribution of muons originating from beauty hadrons is dominant at high $\pT$ ($\pT > 7$~GeV/$c$).
New measurements in central \PbPb{} collisions at \twosevensixnn, with a significantly extended $\pT$ coverage and a higher precision compared to the previous ALICE publication~\cite{Abelev:2012qh}, are reported and compared to the results at \fivenn. The computation of the $\RAA$ makes use of the measured pp references published in~\cite{Abelev:2012qh,Acharya:2019mky}.
Detailed comparisons with model calculations with different implementations of in-medium energy loss are discussed as well.

\section{Experimental apparatus and data samples}\label{sec:detdata}

The ALICE apparatus and its performance are described in~\cite{Aamodt:2008zz,Abelev:2014ffa}.
The analysis is based on the detection of muons in the forward muon spectrometer covering the pseudorapidity interval $-4 < \eta < -2.5$. Note that the muon spectrometer covers a negative $\eta$ range in the ALICE reference frame and consequently a negative $y$ range.
The results are chosen to be presented with a positive $y$ notation, due to the symmetry of the collision system.
The muon spectrometer consists of a front absorber of $10$ nuclear interaction lengths ($\lambda_{\rm I}$) filtering hadrons, followed by five tracking stations, each composed of two planes of Cathode Pad Chambers, with the third station inside a dipole magnet with a field integral of $3$~T$\times$m.
The tracking system is complemented with two trigger stations, each equipped with two planes of Resistive Plate Chambers downstream an iron wall of $7~\lambda_{\rm I}$.
Finally, a conical absorber shields the muon spectrometer against secondary particles produced by the interaction of primary particles at large $\eta$ in the beam pipe.
The Silicon Pixel Detector (SPD), made of two cylindrical layers covering the pseudorapidity intervals $\abs{\eta} < 2$ and $\abs{\eta} < 1.4$, is employed for the reconstruction of the primary vertex.
Two V0-scintillator arrays covering $2.8 < \eta < 5.1$ and $-3.7 < \eta < -1.7$ provide a minimum bias (MB) trigger defined as the coincidence of signals from the two hodoscopes.
The V0 detectors are also used to classify events according to their centrality, determined from a fit of the total signal amplitude based on a two-component particle production model connected to the collision geometry using the Glauber formalism~\cite{Adam:2015ptt}.
The centrality intervals are defined as percentiles of the \PbPb{} hadronic cross section.
The V0 and the Zero Degree Calorimeters (ZDC), placed at $\pm 112.5$~m from the interaction point along the beam direction, are used for the event selection.

The results presented in this letter are based on the data sample recorded with the ALICE detector during the $2015$ \PbPb{} run at a centre-of-mass energy \fivenn.
For the comparison with measurements at lower energy, \twosevensixnn, the $2011$ data sample is used in order to extend the $\pT$ coverage with respect to the published results from the $2010$ data sample~\cite{Abelev:2012qh}.
The analysis of the two data samples is based on muon-triggered events requiring a MB trigger and at least one track segment in the muon trigger system with a $\pT$ larger than a programmable threshold~\cite{Aamodt:2008zz}.
Data were collected with two $\pT$-trigger thresholds of about $1$ ($0.5$) and $4.2$ ($4.2$)~\GeVc at \fivenn (\twosevensixnn).
The $\pT$ threshold of the trigger algorithm is set such that the corresponding efficiency for muon tracks is $50\%$.
In the following, the low- and high-$\pT$ trigger-threshold samples are referred to as MSL and MSH, respectively.
The beam-induced background is reduced offline using the V0 and ZDC timing information, and electromagnetic interactions are removed by requiring a minimum energy deposited in the ZDC~\cite{ALICE:2012aa,Cortese:2019nnv}.
Only events with a primary vertex within $\pm 10$~cm along the beam line are analysed.
Finally, the measurements are done in the three representative centrality classes 0--10\%, 20--40\% and 60--80\% to investigate the evolution of the $\RAA$ with the collision centrality. After the event selection, the data samples correspond to integrated luminosities of about $21.9$ ($224.8$) $\mu$b$^{-1}$ and $4.0$ ($71.0$) $\mu$b$^{-1}$ for MSL- (MSH-) triggered events at $\sNN~=~5.02$ and $2.76$~\TeV, respectively.
The integrated luminosity is derived from the number of muon-triggered events.
These muon-triggered events are normalised by a factor, inversely proportional to the probability of having a muon trigger in a MB event in a given centrality class, calculated from the relative count rate between the muon and MB triggers. 

\section{Analysis procedure} \label{sec:ana}

\subsection{Measurement of muons from heavy-flavour hadron decays}

Standard selection criteria are applied to the muon candidates~\cite{Acharya:2019mky}.
Tracks in the muon spectrometer are reconstructed within the pseudorapidity range $-4 < \eta < -2.5$ and they are required to have a polar angle measured at the exit of the absorber in the interval $170^\circ < \theta_{\rm abs} < 178^\circ$.
Furthermore, tracks are identified as muons if they match a track segment in the trigger system.
Finally, the remaining beam-induced background is reduced by requiring the distance of the track to the primary vertex measured in the transverse plane (DCA, distance of closest approach) weighted with its momentum ($p$), $p\times{\rm DCA}$, to be smaller than $6\times\sigma_{p{\rm DCA}}$, where $\sigma_{p{\rm DCA}}$ is the width of the distribution.

The nuclear modification factor $\RAA$ of muons from heavy-flavour hadron decays is measured down to $\pT=3$~\GeVc and up to $\pT=20$~\GeVc in all centrality classes 
at \fivenn{} 
and in the \cent{0}{10} centrality class at \twosevensixnn. The $\RAA$  is computed for $\pT > 3$~GeV/$c$ in order to limit the systematic uncertainty on the subtraction of the background of muons from light-hadron decays, which increases with decreasing $\pT$. These measurements are performed by using MSL-triggered events up to $\pT=7$~\GeVc and MSH-triggered events for $\pT$~$>$~$7$~\GeVc.
In the selected $\pT$ interval, after the selection criteria are implemented, the main background contributions to the muon yields consist of muons from primary charged-pion and kaon decays for $\pT < 6$~GeV/$c$, and muons from W-boson, Z-boson, and $\gamma^\star$ (Drell-Yan process) decays for $\pT > 13$~\GeVc.
Two additional small contributions of muons from secondary (charged) light-hadron decays in the interval $3 < \pT < 5$~\GeVc, resulting from the interaction of light hadrons with the material of the front absorber and of muons from \jpsi{} decays over the entire $\pT$ range, are also considered.
Therefore, the $\pT$-differential $\RAA$ of muons from heavy-flavour hadron decays in a given centrality class is expressed as
%\begin{linenomath}
\begin{equation}
\displaystyle{%
\RAA(\pT, y) = \displaystyle{%
\frac{%
\left(%
\displaystyle\frac{\dd^{2}N^{\mupm}}{\dd\pT\dd y} - \sum_{{\rm non\mbox{-}HF}\to\mupm}
\displaystyle\frac{\dd^{2}N^{{\rm non\mbox{-}HF}\to\mupm}}{\dd\pT\dd y}
\right)_{\rm Pb-Pb}}
{%
\avg{\TAA}\times
\left(%
\displaystyle{\frac{\dd^{2}\sigma^{{\rm c, b}\rightarrow\mupm}}{\dd\pT\dd y}}\right)_{\rm pp}},}}
\label{eq:RAAdet}
\end{equation}
%\end{linenomath}
where $\dd^{2}N^{\mupm} / \dd\pT\dd y$ is the differential yield of inclusive muons and $\sum_{{\rm non\mbox{-}HF}\rightarrow\mupm}\dd^{2}N^{{\rm non\mbox{-}HF}\rightarrow\mupm} / \dd\pT\dd y$ refers to the differential yields of muons from various non heavy-flavour sources in Pb--Pb collisions, as indicated above Eq.~(\ref{eq:RAAdet}).
In the denominator, $\dd^{2}\sigma^{{\rm c, b}\rightarrow\mupm} / \dd\pT\dd y$ is the pp differential production cross section of muons from heavy-flavour hadron decays at the same centre-of-mass energy and in the same kinematic region (see \cite{Acharya:2019mky,Abelev:2012qh}) as in \PbPb{} collisions.

\subsection{Pb--Pb collisions at \fivenn{}}

\subsubsection{Efficiency corrections}

The inclusive muon yields in Pb--Pb collisions at \fivenn{} are corrected for detector acceptance and detection efficiencies ($A\times\epsilon$) using the procedure described in previous publications~\cite{Abelev:2012qh,Acharya:2019mky}.
In peripheral collisions, $A\times\epsilon$ amounts to about $90\%$ with almost no $\pT$ dependence in the region of interest for MSL-triggered events, while for MSH-triggered events the $A\times\epsilon$ increases with $\pT$ from 75\% at $\pT$ = 7~GeV/$c$ towards a plateau at a value close to $90\%$ for $\pT > 14$~\GeVc.
The dependence of the trigger and tracking efficiency on the detector occupancy is determined by embedding simulated muons from heavy-flavour hadron decays in measured MB Pb--Pb events.
A decrease in the efficiency of $6\%$ from peripheral (\cent{60}{80}) to central (\cent{0}{10}) collisions, independent of $\pT$ is observed.

\subsubsection{Estimation of the muon background sources}

The estimation of the contribution of muons from primary $\pipm$ and $\Kpm$ decays is based on a data-tuned Monte Carlo cocktail.
The procedure uses the midrapidity ($\abs{\eta} < 0.8$) $\pipm$ and $\Kpm$ spectra measured by the ALICE collaboration up to $\pT = 20$~\GeVc~\cite{Acharya:2019yoi} in pp and Pb--Pb collisions at \fivenn.
They are further extrapolated to higher $\pT$, up to $\pT = 40$~\GeVc, by means of a power-law fit to extend the $\pT$ coverage to the $\pT$ interval relevant for the estimation of the decay muons up to $\pT = 20$~\GeVc.
Then, the extrapolation to forward rapidities is performed assuming the same suppression of primary $\pipm$ and $\Kpm$ yields from midrapidity up to $y = 4$ according to
\begin{equation}
\bigg\lbrack \frac{\dd^{2}N^{\pipm(\Kpm)}}{\dd\pT\dd y} \bigg\rbrack_{\rm AA} =
\avg{\Ncoll}\times
\bigg\lbrack \RAA^{\pipm(\Kpm)}\bigg\rbrack^{\rm mid-y} \times
\lbrack F^{\pipm(\Kpm)}_{\rm extrap}(\pT, y) \rbrack_{\rm pp} \times
\bigg\lbrack \frac {\dd^{2}N^{\pipm(\Kpm)}}{\dd\pT\dd y} \bigg\rbrack^{{\rm mid}-y}_{\rm pp}.
\label{eq:RAAextrap1}
\end{equation}
Equation~(\ref{eq:RAAextrap1}) can be also expressed as
\begin{equation}
\bigg\lbrack \frac{\dd^{2}N^{\pipm (\Kpm)}}{\dd\pT\dd y} \bigg\rbrack_{\rm AA} =
\lbrack F^{\pipm(\Kpm)}_{\rm extrap}(\pT, y) \rbrack_{\rm pp} \times
\bigg\lbrack \frac{\dd^{2}N^{\pipm(\Kpm)}}{\dd\pT\dd y} \bigg\rbrack^{{\rm mid}-y}_{\rm AA},
\label{eq:RAAextrap2}
\end{equation}
where $\lbrack F^{\pipm(\Kpm)}_{\rm extrap}(\pT, y) \rbrack_{\rm pp}$ is the $\pT$- and $y$-dependent extrapolation factor in pp collisions at $\sqrt s~=~5.02$ TeV, discussed in~\cite{Acharya:2019mky}, which is based on Monte Carlo simulations.
The systematic uncertainty due to the unknown suppression at forward rapidity will be discussed below.
The PYTHIA~$6.4$~\cite{Sjostrand:2006za} and PHOJET~\cite{Engel:1995sb} event generators are employed for the rapidity extrapolation, while PYTHIA~$8.2$ simulations~\cite{Sjostrand:2014zea} with various colour reconnection (CR) options are performed to take into account the rapidity dependence of the $\pT$ extrapolation and its uncertainty.
The $\pT$ and $y$ distributions of muons from primary $\pipm$ and $\Kpm$ decays in Pb--Pb collisions are generated according to a fast detector simulation of the decay kinematics and of the effect of the front absorber~\cite{Acharya:2019mky} using as input the extrapolated $\pipm$ and $\Kpm$ spectra.
For each centrality class, the yields are further subtracted from the inclusive muon distribution.
The total contribution of muons from primary $\pipm$ and $\Kpm$ decays decreases with increasing $\pT$ from about $21\%$ ($13\%$) at $\pT = 3$~\GeVc down to about $7\%$ ($4\%$) at $\pT = 20$~\GeVc in the \cent{60}{80} (\cent{0}{10}) centrality class, with a weak $\pT$ dependence for $\pT > 10$~\GeVc.

The estimation of the background muons from secondary $\pipm$ and $\Kpm$ decays produced in the front absorber is based on Monte Carlo simulations using the HIJING event generator~\cite{Wang:1991hta} and the GEANT$3$ transport package~\cite{Brun:1994aa}.
These simulation results indicate that in the $\pT$ interval of interest, the relative contribution of secondary muons with respect to muons from primary $\pipm$ and $\Kpm$ decays is about $9\%$, independently of both $\pT$ and the collision centrality.
Given the estimated contamination of muons from primary $\pipm$ and $\Kpm$ decays, the contribution of these secondary muons relative to the total muon yield  decreases with increasing $\pT$ from about $2\%$ ($1\%$) at $\pT = 3$~\GeVc in the \cent{60}{80} (\cent{0}{10}) centrality class to less than $1\%$ at $\pT = 5$~\GeVc for all centrality classes.

The estimation of the contribution of muons from W-boson decays and dimuons from Z-boson and $\gamma^{\star}$ decays, which is relevant in
the high-$\pT$ region, is based on the POWHEG NLO event generator~\cite{Alioli:2008gx} combined with PYTHIA $6.4.25$~\cite{Sjostrand:2006za} for the parton shower, which reproduces within uncertainties the W- and Z-boson production in various LHC experiments~\cite{Chatrchyan:2011ua,Chatrchyan:2014csa,Aad:2014bha,Khachatryan:2015pzs,Alice:2016wka}.
These simulations include the CT$10$ PDF set~\cite{Lai:2010vv} and the EPS$09$ NLO parameterisation~\cite{Eskola:2009uj} of the nuclear modification of the PDFs. In order to account for isospin effects, muons from W-boson decays and dimuons from Z-boson decays and $\gamma^{\star}$ decays are simulated separately in pp, np, pn, and nn collisions. 
A weighted sum of the production cross sections in the four systems is performed to obtain the production cross section per nucleon--nucleon collision for the Pb--Pb system.
The latter is further scaled with $\avg{\TAA}$ in a given centrality class in order to estimate the corresponding relative contribution of W and Z/$\gamma^{\star}$ with respect to inclusive muons.
The relative contribution of muons from W and Z/$\gamma^{\star}$ with respect to inclusive muons is negligible for $\pT < 13$~\GeVc and it increases with $\pT$ and the collision centrality from about $3\%$ ($6\%$) at $\pT = 14$~\GeVc up to $18\%$ ($36\%$) at $\pT = 20$~\GeVc in the \cent{60}{80} (\cent{0}{10}) centrality class.

The contribution of muons from \jpsi{} decays is estimated by extrapolating the \jpsi{} $\pT$ and $y$ spectra measured by ALICE at forward rapidity ($2.5 < y < 4$) in the interval of $\pT < 12$~\GeVc~\cite{Adam:2016rdg}.
The \jpsi{} $\pT$ and rapidity spectra are extrapolated by means of a power-law and Gaussian function up to $\pT = 50$~\GeVc and $\abs{y} = 6.5$, respectively.
Then, the decay muon distributions are estimated with a fast detector simulation using the extrapolated \jpsi{} distributions as inputs, similar to pp collisions~\cite{Acharya:2019mky}.
In the $10\%$ most central collisions, the relative contribution to the inclusive muon distribution varies between $0.5$ and $4\%$, with the maximum fraction at intermediate $\pT$ ($ 4 < \pT < 6$~\GeVc).

\subsubsection{Systematic uncertainties}

The systematic uncertainties of the $\RAA$ of muons from heavy-flavour hadron decays at \fivenn{} are evaluated considering the following sources: uncertainties of the inclusive muon yields and background contributions in Pb--Pb collisions, the pp reference, and the normalisation in both pp and \PbPb{} collisions.

The procedure to determine the systematic uncertainty on the inclusive muon yields  is similar to that described in~\cite{Acharya:2019mky} and includes the following contributions:
i) the muon tracking efficiency ($1.5\%$),
ii) the muon trigger efficiency resulting from the intrinsic efficiency of the muon trigger chambers and the response of the trigger algorithm ($1.4\%$ ($3\%$) for the MSL (MSH) data sample), and iii) the choice of the $\chi^{2}$ selection used in defining the matching of tracks reconstructed in the tracking system with those in the trigger system 
($0.5\%$). 
These systematic uncertainties are approximately independent of centrality and $\pT$ in the region of interest.
The systematic uncertainty arising from the dependence of $A\times\epsilon$ on the detector occupancy, obtained from a fit with a constant of the $\pT$-differential ratio of the efficiency in a given centrality class to that in peripheral collisions, increases up to $0.5\%$ when going from peripheral to central collisions. Finally, the systematic uncertainty due to the tracking chamber resolution and alignment is based on a Monte Carlo simulation modeling the tracker response with a parameterisation of the tracking chamber resolution and misalignment effects, as described in~\cite{Alice:2016wka,Acharya:2019mky}.
This systematic uncertainty is negligible for $\pT < 7$~\GeVc and increases up to $12\%$ in the interval $18 < \pT < 20$~\GeVc.

The estimation of the yields of muons from primary $\pipm$ and $\Kpm$ decays is subject to systematic uncertainties arising, as described in~\cite{Acharya:2019mky}, from
i) the uncertainties of the measured midrapidity spectra of $\pipm$ ($\Kpm$) and their $\pT$ extrapolation, which increase from about $3\%$ ($6\%$) to $6\%$ ($13\%$),
ii) the rapidity extrapolation which results in a systematic uncertainty of about $8.5\%$ ($6\%$) for muons from $\pipm$ ($\Kpm$) decays obtained by comparing the results with PYTHIA~$6$ and PHOJET generators,
iii) the rapidity dependence of the $\pT$ extrapolation with a systematic uncertainty, obtained from the PYTHIA~$8$ generator with different CR options, increasing up to about $4\%$ ($2\%$) at $\pT = 20$~\GeVc for $\pipm$ ($\Kpm$), and
iv) the simulation of hadronic interactions in the absorber which leads to a systematic uncertainty of $4\%$ independently of the muon origin, as reported in~\cite{Acharya:2019mky}.
Adding in quadrature the uncertainties coming from each source, the total systematic uncertainty ranges from about $9\%$ ($10\%$) to $13\%$ ($15\%$) as a function of the $\pT$ of muons from primary $\pipm$ ($\Kpm$) decays.
Finally, there is a contribution related to the assumption on the rapidity dependence of the suppression of $\pipm$ and $\Kpm$.
Based on ATLAS measurements in \PbPb{} collisions at \twosevensixnn, which indicate no significant $\eta$ dependence of the charged-particle $\RAA$ up to $\abs{\eta} < 2$~\cite{Aad:2015wga}, the suppression of $\pipm$ and $\Kpm$ is considered to be independent of rapidity up to $y = 4$, and the $\RAA$ of $\pipm$ and $\Kpm$ is varied conservatively within $\pm50\%$.
This uncertainty is propagated to the decay muons and the difference between the upper and lower limits is further divided by $\sqrt{12}$, corresponding to the RMS of a uniform distribution.
Furthermore, the effect of the transport code is conservatively evaluated by varying the estimated yield of muons from secondary $\pipm$ and $\Kpm$ decays by $\pm 100\%$ and dividing also the difference between lower and upper limits by $\sqrt{12}$.

The systematic uncertainty of the extracted muon yields from W and Z/$\gamma^{\star}$ decays is obtained considering the CT$10$ PDF uncertainty~\cite{Lai:2010vv} and a different nuclear modification of the PDF (EKS$98$~\cite{Eskola:2003cc,Eskola:1998df,Mangano:815037} was used as well).
It amounts to $5.9\%$ ($13.2\%$) for muons from W (Z/$\gamma^{\star}$) decays.

The systematic uncertainty of the estimated yields of muons from \jpsi{} decays reflects the uncertainty of the measured \jpsi{} spectra at forward rapidity and their extrapolation to a wider kinematic region.
It varies from about $9\%$ at $\pT = 3$~\GeVc to $34\%$ at $\pT = 20$~\GeVc in central collisions.

Two sources contribute to the systematic uncertainty on the normalisation, the systematic uncertainty of $\avg{\TAA}$ values~\cite{ALICE-PUBLIC-2018-011} 
and  the systematic uncertainty of the normalisation factor needed to calculate the number of equivalent MB events in the muon samples. 
The latter is evaluated comparing the values from the nominal procedure (see section~\ref{sec:detdata}) with those calculated by applying the muon-trigger condition in the analysis of MB events~\cite{Acharya:2019mky}.

The sources of systematic uncertainty affecting the measurement of the pp reference production cross section were evaluated in~\cite{Acharya:2019mky}.
The total systematic uncertainty ranges from $2.1\%$ to $15.1\%$, depending on $\pT$.
A global pp normalisation uncertainty of $2.1\%$, discussed in~\cite{Acharya:2019mky}, is considered as well.
When computing the nuclear modification factor, the systematic uncertainty on track resolution and misalignment is considered to be partially correlated between the pp and Pb--Pb measurements because the pp data were collected just before the Pb--Pb run at \fivenn and the detector conditions remained unchanged.
The other sources of systematic uncertainties are treated as uncorrelated.
The systematic uncertainty on the $\pT$-differential production cross section in pp collisions without including the correlated part of the uncertainty varies from $2.1\%$ to $4.2\%$.
The uncorrelated part of the uncertainty on track resolution and misalignment is due to the different shapes of the $\pT$ distribution between pp and Pb--Pb collisions.
It is estimated by comparing the results with and without correcting the residual misalignment between data and Monte Carlo when calculating the $\RAA$, as detailed in~\cite{Acharya:2019mky}.

The various systematic uncertainties are propagated to the measurement of the yields or nuclear modification factors of muons from heavy-flavour hadron decays and added in quadrature, except for the systematic uncertainties on normalisation which are shown separately.

Table~\ref{tab:SystUnc} presents a summary of the relative systematic uncertainties assigned to the $\pT$-differential yields of muons from heavy-flavour hadron decays in Pb--Pb collisions.
The systematic uncertainty on the pp reference, needed for the computation of the $\RAA$, is also reported.

\begin{sidewaystable}
\caption{Summary of the relative systematic uncertainties of the $\pT$-differential yields of muons from heavy-flavour hadron decays at forward rapidity ($2.5 < y <4$) in \PbPb{} collisions at \fivenn (second and third columns) and $2.76$~\TeV (fourth column).
The systematic uncertainties of the pp reference are also summarised.
For the $\pT$-dependent uncertainties, the minimum and maximum values are reported and correspond to the lowest and highest $\pT$ interval with the exception of the background of muons from light-hadron decays and the $\RAA^{\pipm(\Kpm)}(y)$ assumption, where this is the opposite.
See the text for details.}

\centering
\begin{tabular}{|l|c|c|c|}
\hline
Source & \multicolumn{2}{c|} {\fivenn} & \twosevensixnn \\
& \cent{0}{10} centrality class & \cent{60}{80} centrality class & \cent{0}{10} centrality class \\
\hline
Tracking efficiency      & $1.5\%$ & $1.5\%$ & $2.5\%$ \\
Trigger efficiency       & $1.4\%$ (MSL), $3\%$ (MSH) & $1.4\%$ (MSL), $3\%$ (MSH)& $1.4\%$ (MSL), $2.3\%$ (MSH)\\
Matching efficiency      & $0.5\%$ & $0.5\%$ & $0.5\%$ \\
$A\times\epsilon$        & $0.5\%$ & $0$ &  $1\%$ \\
Resolution and alignment & \cent{0}{12} (\cent{0}{4.1} on $\RAA$) & \cent{0}{12} (\cent{0}{4.1} on $\RAA$)& $1\% \times\pT$ ($\pT$ in~\GeVc) \\
Background subtraction $\mu\leftarrow\pi$ & $<1.6\%$ & $<2.5\%$ & $<1.8\%$ \\
Background subtraction $\mu\leftarrow{\rm K}$ & $<1.6\%$ & $<2.5\%$ & $<4\%$ \\
$\RAA^{\pipm(\Kpm)}(y)$ assumption & \cent{1.3}{4.8} & \cent{1.5}{7.8} & \cent{1.8}{5.2} \\
Background subtraction $\mu\leftarrow{\rm sec.}\pi{\rm /K}$ & \cent{0}{0.8} & \cent{0}{1.4} & \cent{0}{0.9} \\
Background subtraction $\mu\leftarrow{\rm W/Z/}\gamma^{\star}$ & \cent{0}{1.6} & \cent{0}{0.7} & \cent{0}{3.1} \\
Background subtraction $\mu\leftarrow\Jpsi$ & $< 0.4\%$ & $< 0.4\%$ & $< 0.3\%$ \\
Normalisation factor & $0.3\%$ (MSL), $0.7\%$ (MSH) & $0.3\%$ (MSL), $0.7\%$ (MSH) & $0.4\%$ (MSL), $1.6\%$ (MSH)\\
$\avg{\TAA}$ & $0.7\%$ & $2.5\%$ & $0.9\%$ \\
pp reference for $\RAA$ & \cent{2.1}{4.2} & \cent{2.1}{4.2} & \cent{15}{18} ($3 < \pT < 10$~\GeVc data) \\
& & & \cent{30}{34} ($10 < \pT < 20$~\GeVc extrapolation) \\
pp reference (global) for $\RAA$  & $2.1\%$ & $2.1\%$ & $1.9\%$ \\
\hline
\end{tabular}
\label{tab:SystUnc}
\end{sidewaystable}

\subsection{Pb--Pb collisions at \twosevensixnn}

For a direct comparison with lower energy measurements in the same $\pT$ interval, the Pb--Pb data sample at \twosevensixnn, collected in $2011$, was analysed in order to significantly extend the $\pT$ interval of the published $\RAA$ measurements of muons from heavy-flavour hadron decays, which was limited to $4 < \pT < 10$~\GeVc~\cite{Abelev:2012qh}.
Such an improvement is possible due to the larger integrated luminosity ($4\ \mu {\rm b^{-1}}$ and $71\ \mu {\rm b^{-1}}$ for MSL- and MSH-triggered collisions compared to $2.7\ \mu {\rm b^{-1}}$) and the use of a high-$\pT$ muon trigger.

The strategy to extract the yields of muons from heavy-flavour hadron decays in Pb--Pb collisions at \twosevensixnn is similar to that just discussed for \fivenn. Compared to the latter case, the $A\times\epsilon$ exhibits the same trend as a function of $\pT$, although the values are smaller due to the status of the tracking chambers (larger number of inactive channels).
The factor $A\times\epsilon$ saturates at a value close to $80\%$ in the high-$\pT$ region for peripheral collisions (\cent{60}{80} centrality class).
%The dependence of the  efficiency on the detector occupancy, hence on the collision centrality, is investigated using the embedding procedure as at \fivenn.
A decrease of the efficiency of $4\%$ from peripheral collisions to the $10\%$ most central collisions, due to the detector occupancy, is seen.
The fractions of the various background sources with respect to the inclusive muon yields at \twosevensixnn are compatible with the ones measured at \fivenn.
The fraction of muons from primary $\pipm$ and $\Kpm$ decays with respect to inclusive muons varies between about $3\%$ and $14\%$ in the \cent{0}{10} centrality class, the largest values being obtained at $\pT = 3$~\GeVc.
On the other hand, the fraction of muons from secondary $\pipm$ and $\Kpm$ decays reaches about $1\%$ at $\pT = 3$~\GeVc.
The fraction of muons from electroweak-boson decays is significant at high $\pT$, where it reaches about $30\%$ in the interval $16.5 < \pT < 20$~\GeVc for central collisions.
Finally, the component of muons from \jpsi{} decays is small over the whole $\pT$ interval with a maximum of $4\%$ at intermediate $\pT$ ($\sim 6$~\GeVc) in central collisions.
The same sources of systematic uncertainties as for the $\sNN$ = 5.02 TeV analysis are considered and same methods to estimate them are employed, except the systematic uncertainty of the tracking chamber resolution and alignment which varies linearly with $\pT$ as $1\%\times \pT$ ($\pT$~in \GeVc)~\cite{Abelev:2012qh}.
The $\pT$-differential cross section of muons from heavy-flavour hadron decays in pp collisions at \twosevensix measured in the intervals $2.5 < y < 4$ and $ 3 < \pT < 10$~\GeVc is used for the $\RAA$ computation~\cite{Abelev:2012qh}.
The measured production cross section is extrapolated up to $\pT = 20$~\GeVc using fixed-order plus next-to-leading logarithms (FONLL) calculations~\cite{Cacciari:1998it,Cacciari:2012ny}.
The systematic uncertainty of the $\pT$-differential production cross section in pp collisions at \twosevensix{} varies within \cent{15}{18} in $3 < \pT < 10$~\GeVc.
At higher $\pT$, the systematic uncertainty, which also includes the systematic uncertainty on the FONLL calculations, reaches~\cent{30}{34}.

A summary of all systematic uncertainties taken into account in the measurement of the $\pT$-differential yields of muons from heavy-flavour hadron decays at \twosevensixnn is reported in Table~\ref{tab:SystUnc}, including the uncertainties of the pp reference.

\section{Results and model comparisons}\label{sec:res}

\begin{figure}[!hbt]
\begin{center}
\includegraphics[width=0.45\textwidth]{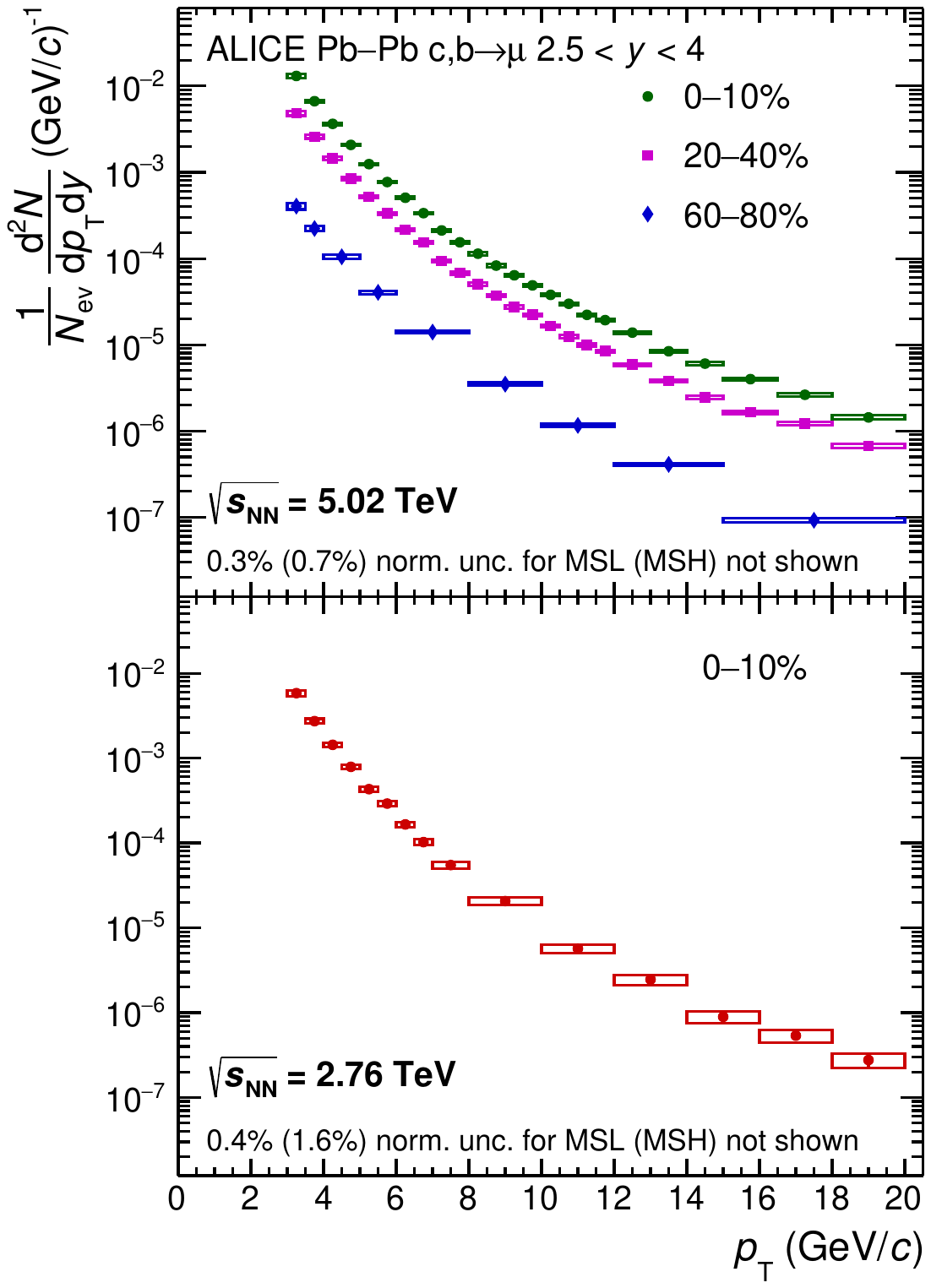}
\caption{The $\pT$-differential yields of muons from heavy-flavour hadron decays at forward rapidity ($2.5 < y < 4$) in central (\cent{0}{10}), semi-central (\cent{20}{40}), and peripheral (\cent{60}{80}) Pb--Pb collisions at \fivenn{} (upper panel), and in central (\cent{0}{10}) Pb--Pb collisions at \twosevensixnn (lower panel).
Statistical uncertainties (vertical bars) and systematic uncertainties (open boxes) are shown.
The additional systematic uncertainty on normalisation in Pb--Pb collisions at $\sNN$~$=$~$5.02$ ($2.76$)~\TeV for MSL- and MSH-triggered events, respectively, is not included in the uncertainty boxes. (see Table~\ref{tab:SystUnc}).}
\label{Fig:pTyield}
\end{center}
\end{figure}

The $\pT$-differential yields of muons from heavy-flavour hadron decays normalised to the equivalent number of MB events at forward rapidity ($2.5 < y < 4$) in central, semi-central and peripheral Pb--Pb collisions at \fivenn are shown in Fig.~\ref{Fig:pTyield} (upper panel).
The same observable measured in central Pb--Pb collisions at \twosevensixnn is displayed in the lower panel of Fig.~\ref{Fig:pTyield}.
The measurements are performed over a wide $\pT$ range from 3 to 20 GeV/$c$ for all centrality classes.

\begin{figure}[!thb]
\begin{center}
\includegraphics[width=0.8\textwidth]{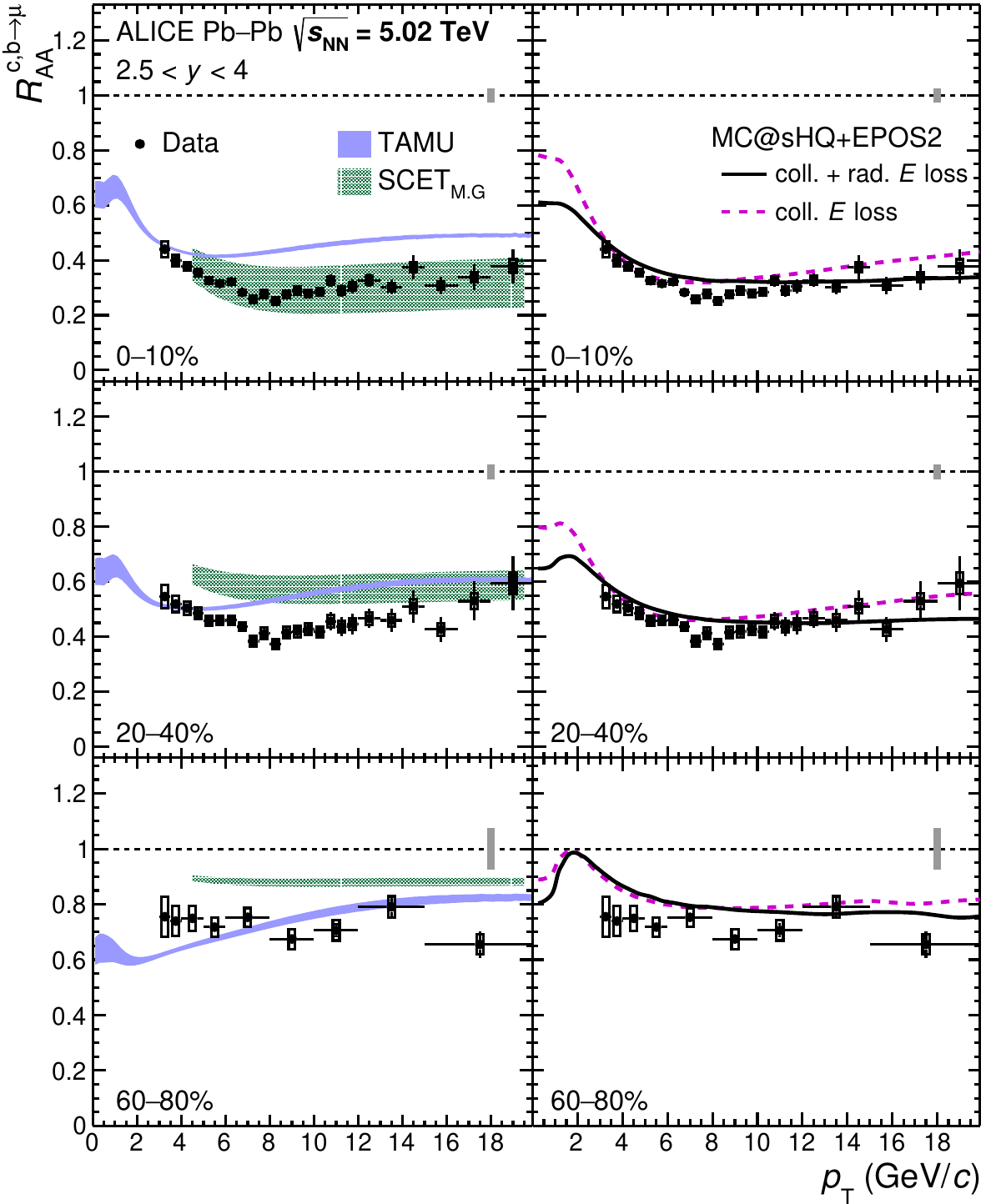}
\caption{The $\pT$-differential nuclear modification factor $\RAA$ of muons from heavy-flavour hadron decays at forward rapidity ($2.5 < y < 4$) in central (\cent{0}{10}, top), semi-central (\cent{20}{40}, middle), and peripheral (\cent{60}{80}, bottom) Pb--Pb collisions at \fivenn{} (symbols).
Statistical (vertical bars) and systematic uncertainties (open boxes) are shown.
The filled boxes centered at $\RAA = 1$ represent the normalisation uncertainty of pp and Pb--Pb measurements.
Horizontal bars reflect the bin widths and the values are shown at the centre of the bin.
Left: the measured $\RAA$ is compared with the TAMU and SCET models~\cite{He:2014cla,Kang:2016ofv} displayed with their uncertainty bands.
Right: the measured $\RAA$ is compared with MC@sHQ+EPOS2 model calculations with pure collisional energy loss (full lines) and a combination of collisional and radiative energy loss (dashed lines)~\cite{Nahrgang:2013xaa,Nahrgang:2013saa}.}
\label{Fig:pTRAA}
\end{center}
\end{figure}

The $\pT$-differential $\RAA$ of muons from heavy-flavour hadron decays at forward rapidity ($2.5 < y < 4$) in Pb--Pb collisions at \fivenn is presented in Fig.~\ref{Fig:pTRAA} for the same centrality classes as in Fig.~\ref{Fig:pTyield}.
An increasing reduction of the yield of muons from heavy-flavour hadron decays with increasing centrality with respect to the pp reference scaled by the average nuclear overlap function is clearly seen.
The suppression is largest at intermediate $\pT$, in the interval from about 6 to 10~\GeVc, and reaches a factor of about three in the $10\%$ most central collisions. Such behaviour is more pronounced in central and semi-central collisions, while moving towards peripheral collisions, the suppression presents no significant $\pT$ dependence.     
In minimum bias p--Pb~collisions at \fivenn, where the formation of an extended QGP is not expected, the nuclear modification factor $\RpPb$ of muons from heavy-flavour hadron decays is consistent with unity at $\pT>6$~\GeVc~\cite{Acharya:2017hdv}.
The latter measurement confirms that the strong suppression observed in Pb--Pb collisions results from final-state interactions of charm and beauty quarks with the QGP.
The evolution of $\RAA$ as a function of centrality is compatible with the dependence of the heavy-quark energy loss on the medium density and the average path length in the medium, both of which are larger in central than in peripheral collisions.

The measured $\RAA$ is compared with various model predictions such as TAMU~\cite{He:2014cla} and SCET~\cite{Kang:2016ofv} (Fig.~\ref{Fig:pTRAA}, left), and MC@sHQ+EPOS2~\cite{Nahrgang:2013xaa,Nahrgang:2013saa} (Fig.~\ref{Fig:pTRAA}, right).
In the TAMU model, the interactions are described by elastic collisions only.
The perturbative QCD (pQCD)-based SCET model implements medium-induced gluon radiation via modified splitting functions with finite quark masses.
These SCET calculations depend on the coupling constant $g$ which describes the coupling strength between hard partons and the QGP medium.
Its value is $g = 1.9$--$2$.
In the MC@sHQ+EPOS2 model, two different options are considered, energy loss from medium-induced gluon radiation and collisional (elastic) processes or only collisional energy loss.
In the scenario with pure collisional energy loss, the scattering rates are scaled by a global factor $K$ larger than unity ($K = 1.5$) in order to reproduce the $\RAA$ and elliptic flow of open heavy-flavour hadrons measured at midrapidity at the LHC~\cite{Nahrgang:2013xaa}.
With a combination of collisional and radiative energy loss, the scaling factor is $K = 0.8$.
All these models also consider a nuclear modification of the PDF (EPS09)~\cite{Eskola:2009uj}.
Note that in the MC@sHQ+EPOS2 model shadowing is not considered for beauty-quark production.
In addition to independent fragmentation, a contribution of hadronisation via quark recombination is included in all models with the exception of SCET.
The SCET model is based on pQCD calculations of high-$\pT$ parton energy loss and provides a fair description of the data in central collisions, but it deviates from the data in non-central collisions.
The TAMU calculations, which do not include radiative energy loss processes, underestimate the suppression at $p_{\rm T} > 6$~GeV/$c$ in central and semi-central collisions, in particular.
Both versions of the MC@sHQ+EPOS2 model, without and with radiative energy loss, describe the measurement within uncertainties for all centrality classes over the entire $\pT$ interval.

The results obtained at forward rapidity for muons from heavy-flavour hadron decays at $\sNN$~$=$~$5.02$~TeV complement those obtained at midrapidity for the electrons from heavy-flavour hadron decays~\cite{Acharya:2019mom} by the ALICE collaboration as well as the prompt D-meson~\cite{Acharya:2018hre,Sirunyan:2017xss} and beauty measurements via $\Bpm$ mesons~\cite{Sirunyan:2018ktu}, non-prompt $\Dzero$~\cite{Sirunyan:2018ktu} and \jpsi{}~\cite{Sirunyan:2017isk} by the ALICE and CMS collaborations.
The measured $\RAA$ of muons from heavy-flavour hadron decays for $\pT > 8$~\GeVc is compatible with that obtained for beauty ($\Dzero$ and \jpsi{} from beauty hadrons, $\Bpm$) for $\pT^{\rm hadron} > 10$~\GeVc~\cite{Sirunyan:2017isk,Sirunyan:2018ktu} within uncertainties, although in a different kinematic region (different $\pT$ and $y$ intervals).

\begin{figure}[!tbh]
\begin{center}
\includegraphics[width=.65\textwidth]{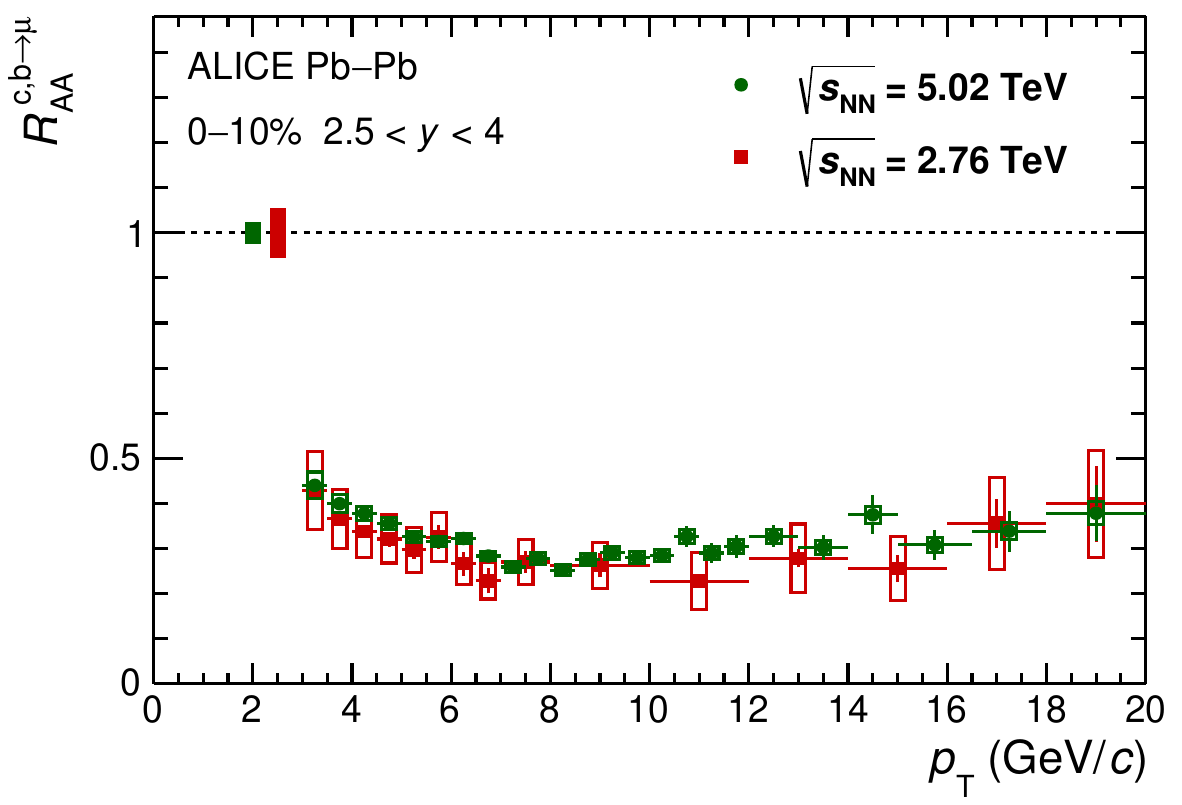}
\caption{Comparison of the $\pT$-differential nuclear modification factor of muons from heavy-flavour hadron decays at forward rapidity ($2.5 < y < 4$) in central Pb--Pb collisions at \fivenn (green symbols) and \twosevensixnn (red symbols).
Statistical (vertical bars) and systematic uncertainties (open boxes) are shown.
The filled boxes centered at $\RAA = 1$ are the normalisation uncertainties. Horizontal bars represent the bin widths.}
\label{Fig:RAAComp}
\end{center}
\end{figure}

A comparison of the $\RAA$ of muons from heavy-flavour hadron decays in the $10\%$ most central Pb--Pb collisions at $\sNN$~$=$~$2.76$ and $5.02$~\TeV is presented in Fig.~\ref{Fig:RAAComp}.
The comparison illustrates the improvement of the precision of the  measurement at \fivenn with respect to that at \twosevensixnn.
The total systematic uncertainty on the $\RAA$ at \fivenn is reduced by a factor of about $3$ to $6$, depending on $\pT$, compared to the same measurement at \twosevensixnn using the $2011$ data sample.
The reasons for such an improvement are twofold.
The detector conditions were more stable during the \fivenn than the \twosevensixnn data taking campaign and therefore better described in the simulations.
Moreover, as the pp data at \five were collected just a few days before the Pb--Pb run at \fivenn, the detector conditions were comparable and the systematic uncertainty on alignment and resolution between the two systems partially cancel when computing $\RAA$, as discussed in section~\ref{sec:ana}.
The present measurement at \twosevensixnn is in agreement with the published results obtained at the same centre-of-mass energy in a smaller $\pT$ interval ($4 < \pT < 10$~GeV/$c$) with larger uncertainties~\cite{Abelev:2012qh}. The precision is increased by a factor $1.1$--$1.6$, mainly due to a better understanding of the detector response and a new data-driven strategy for the estimation of the contribution of muons from primary light-hadron decays.
The comparison between the results obtained at the two centre-of-mass energies indicates that the suppression of heavy quarks at \fivenn is similar to that at \twosevensixnn, as already observed in the midrapidity region for electrons from heavy-flavour hadron decays~\cite{Acharya:2018upq, Acharya:2019mom} and prompt D mesons~\cite{Acharya:2018hre}.
This similarity between the $\RAA$ measurements at the two energies may result from the interplay of the following two effects as discussed in~\cite{Djordjevic:2015hra}: a flattening of the $\pT$ spectra of charm and beauty quarks with increasing collision energy, and a medium temperature estimated to be higher by about $7\%$ at \fivenn than at $2.76$~\TeV.
The former would decrease the heavy-quark suppression (increase the $\RAA$) by about $5\%$ if the medium temperature remains unchanged, while the latter would increase the suppression (decrease the $\RAA$) by about $10\%$ ($5\%$) for charm (beauty) quarks.

The measured $\RAA$ at \twosevensixnn{} is compatible with that measured for muons from heavy-flavour hadron decays in $\abs{\eta} < 1$ with the ATLAS detector~\cite{Aaboud:2018bdg} and for electrons from heavy-flavour hadron decays in the interval $\abs{y} < 0.6-0.8$ by the ALICE collaboration~\cite{Adam:2016khe}. The same behaviour is also observed at \fivenn{} when comparing the $\RAA$ of muons from heavy-flavour hadron decays with that measured at midrapidity for electrons from heavy-flavour hadron decays~\cite{Acharya:2019mom}. This confirms that heavy quarks suffer a strong in-medium energy loss over a wide rapidity interval. The similarity of the suppression in the two rapidity regions does not imply that heavy quarks lose similar energy. The observed trend may also result from the interplay of several effects such as the shape of initial heavy-quark $\pT$ spectra and the path-length dependence of the heavy-quark energy loss, as discussed in~\cite{Prado:2019ste}. Indeed, the properties of the QGP medium differ between mid and forward rapidity. The measured charged-particle multiplicity densities are smaller at forward rapidity than at 
midrapidity~\cite{Adam:2016ddh}. The created medium is also smaller and consequently the traveled path length is shorter at forward rapidity.

\begin{figure}[!tbh]
\begin{center}
\includegraphics[width=.65\textwidth]{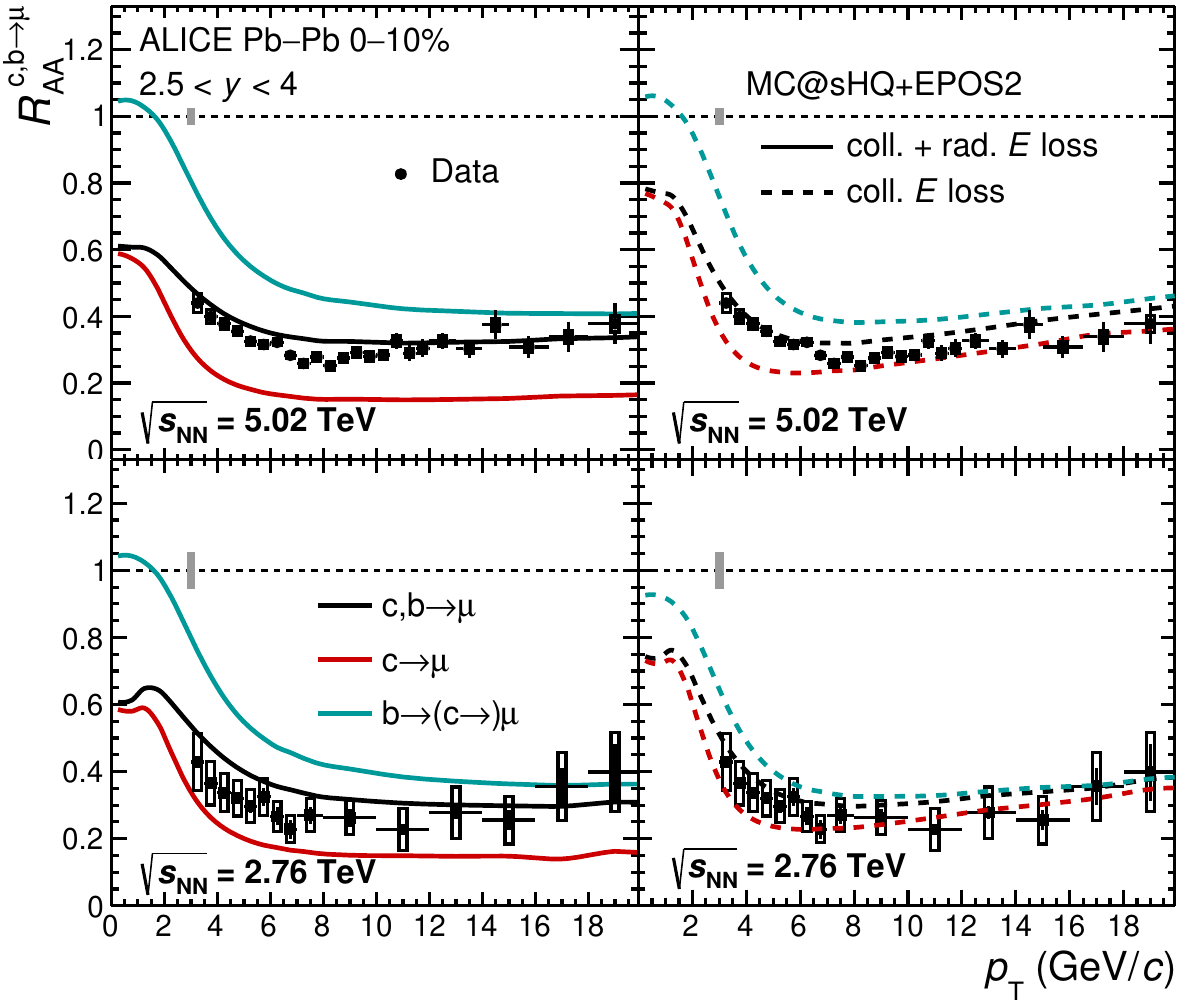}
\caption{Comparison of the $\pT$-differential nuclear modification factors $\RAA$ of muons from heavy-flavour hadron decays at forward rapidity ($2.5 < y < 4$) in central Pb--Pb collisions at \fivenn (top) and \twosevensixnn (bottom) with MC@sHQ+EPOS2 calculations~\cite{Nahrgang:2013xaa,Nahrgang:2013saa} with different scenarios considering either a combination of collisional and radiative energy loss (left) or a pure collisional energy loss (right).
The predictions are shown for muons from heavy-flavour hadron decays, muons from only charm-hadron decays and muons from only beauty-hadron decays.}
\label{Fig:pTRAA0-10comp}
\end{center}
\end{figure}

The $\pT$ distributions of muons from heavy-flavour hadron decays are sensitive to energy loss of both charm and beauty quarks.
Due to the decay kinematics and the charm- and beauty-quark $\pT$-differential production cross sections, one expects that for $\pT \lesssim 5$~GeV/$c$ the distributions are predominantly sensitive to the charm in-medium energy loss.
FONLL calculations~\cite{Cacciari:1998it,Cacciari:2012ny} predict that in pp collisions at \five more than $70\%$ of muons from heavy-flavour hadron decays originate from beauty quarks in the high-$\pT$ region ($\pT > 10$~\GeVc) and this contribution reaches $75\%$ in the interval $18 < \pT < 20$~\GeVc.
Therefore, the strong suppression of muons from heavy-flavour hadron decays in the high-$\pT$ region is expected to be dominated by the in-medium energy loss of beauty quarks.
In order to further interpret the results, Fig.~\ref{Fig:pTRAA0-10comp} shows a comparison with MC@sHQ+EPOS2 predictions for muons from charm- and beauty-hadron decays, separately, and for muons from the combination of the two, in central (\cent{0}{10}) Pb--Pb collisions at \fivenn (top) and $2.76$~\TeV (bottom).
The predictions considering the combination of elastic and radiative energy loss and pure elastic energy loss are shown in the left and right panels, respectively.
Both versions of the MC@sHQ+EPOS2 model provide a fair description of the measured $\RAA$ of muons from heavy-flavour hadron decays in central Pb--Pb collisions at \fivenn within uncertainties.
A similar agreement between data and MC@sHQ+EPOS2 is achieved at \twosevensixnn{} although the model tends to slightly overestimate the measured $\RAA$ at low/intermediate $\pT$.
The measured $\RAA$ at large $\pT$ is closer to the model calculations for muons from beauty-hadron decays than for muons from charm-hadron decays when considering both elastic and radiative energy loss.
For the scenario involving only collisional energy loss, the predicted difference between the suppression of muons from charm and beauty-hadron decays is less pronounced.
The predicted ratio of the $\pT$-differential $\RAA$ of muons from beauty-hadron decays to that of muons from charm-hadron decays for $\pT > 10$~\GeVc is in the range $1.2$--$1.4$ for the scenario involving only collisional energy loss and in the range $2.5$--$2.8$ when considering both elastic and radiative energy loss, depending on $\pT$ and centre-of-mass energy.
It is worth mentioning that the MC@sHQ+EPOS2 model is characterised by a large running coupling constant $\alpha_{\rm s}$ and a reduced Debye mass in the elastic heavy-quark scattering generating the radiation~\cite{Rapp:2018qla}.
As a consequence, the radiative energy loss neglects finite path-length effects due to the gluon formation outside the QGP and is overestimated at high $\pT$.
Such an effect is expected to be more pronounced for charm quarks than for beauty quarks due to the dead-cone effect~\cite{Dokshitzer:2001zm}.

\section{Conclusions}%
\label{sec:concl}

In summary, the $\pT$-differential normalised yield and the nuclear modification factor $\RAA$ of muons from semi-leptonic decays of charm and beauty hadrons was measured at forward rapidity ($2.5 < y < 4$) for the first time over the wide $\pT$ interval $3 < \pT < 20$~\GeVc in central, semi-central, and peripheral Pb--Pb collisions at \fivenn, and in central Pb--Pb collisions at \twosevensixnn with reduced systematic uncertainties compared to previous measurements.

The measured $\RAA$ shows a clear evidence of a strong suppression, up to a factor of three in the 10\% most central collisions with respect to the binary-scaled pp reference, for both collision energies. This suppression pattern is compatible with a large heavy-quark in-medium energy loss.
The strong suppression which persists in the high-$\pT$ region, up to $\pT = 20$~\GeVc, indicates that beauty quarks lose a significant fraction of their energy in the medium.
The suppression becomes weaker from central to peripheral collisions.
The evolution of $\RAA$ with the collision centrality reflects the dependence of energy loss on the path length in the QGP and the QGP energy density. 

The $\RAA$ measurements have the potential to discriminate between different  model calculations. The $\RAA$ is in fair agreement with transport model calculations that consider both collisional and radiative energy loss.
The MC@sHQ+EPOS2 transport model including a hydrodynamic description of the medium, coupled with different implementations of the in-medium parton energy loss, describes the measured $\RAA$ well over the whole $\pT$ interval in central, semi-central, and peripheral collisions within uncertainties. This comparison brings new constraints on the relative in-medium energy loss of charm and beauty quarks.

The suppression is compatible with that measured at central rapidity for electrons from heavy-flavour hadron decays. These new precise $\RAA$ measurements carried out over a wide $\pT$ interval at forward rapidity in Pb--Pb collisions at \fivenn with smaller uncertainties with respect to same measurements at \twosevensixnn, currently only accessible by ALICE in central collisions, bring significant constraints on the modeling of the longitudinal dependence of the open heavy-flavour $\RAA$. Therefore, the obtained results provide further insight on the in-medium parton energy loss mechanisms and, ultimately, will help determining the transport properties of the hot and dense deconfined QCD medium in the full phase space.

\newenvironment{acknowledgement}{\relax}{\relax}
\begin{acknowledgement}
\section*{Acknowledgements}
% Version: 2020-09-14

The ALICE Collaboration would like to thank all its engineers and technicians for their invaluable contributions to the construction of the experiment and the CERN accelerator teams for the outstanding performance of the LHC complex.
The ALICE Collaboration gratefully acknowledges the resources and support provided by all Grid centres and the Worldwide LHC Computing Grid (WLCG) collaboration.
The ALICE Collaboration acknowledges the following funding agencies for their support in building and running the ALICE detector:
A. I. Alikhanyan National Science Laboratory (Yerevan Physics Institute) Foundation (ANSL), State Committee of Science and World Federation of Scientists (WFS), Armenia;
Austrian Academy of Sciences, Austrian Science Fund (FWF): [M 2467-N36] and Nationalstiftung f\"{u}r Forschung, Technologie und Entwicklung, Austria;
Ministry of Communications and High Technologies, National Nuclear Research Center, Azerbaijan;
Conselho Nacional de Desenvolvimento Cient\'{\i}fico e Tecnol\'{o}gico (CNPq), Financiadora de Estudos e Projetos (Finep), Funda\c{c}\~{a}o de Amparo \`{a} Pesquisa do Estado de S\~{a}o Paulo (FAPESP) and Universidade Federal do Rio Grande do Sul (UFRGS), Brazil;
Ministry of Education of China (MOEC) , Ministry of Science \& Technology of China (MSTC) and National Natural Science Foundation of China (NSFC), China;
Ministry of Science and Education and Croatian Science Foundation, Croatia;
Centro de Aplicaciones Tecnol\'{o}gicas y Desarrollo Nuclear (CEADEN), Cubaenerg\'{\i}a, Cuba;
Ministry of Education, Youth and Sports of the Czech Republic, Czech Republic;
The Danish Council for Independent Research | Natural Sciences, the VILLUM FONDEN and Danish National Research Foundation (DNRF), Denmark;
Helsinki Institute of Physics (HIP), Finland;
Commissariat \`{a} l'Energie Atomique (CEA) and Institut National de Physique Nucl\'{e}aire et de Physique des Particules (IN2P3) and Centre National de la Recherche Scientifique (CNRS), France;
Bundesministerium f\"{u}r Bildung und Forschung (BMBF) and GSI Helmholtzzentrum f\"{u}r Schwerionenforschung GmbH, Germany;
General Secretariat for Research and Technology, Ministry of Education, Research and Religions, Greece;
National Research, Development and Innovation Office, Hungary;
Department of Atomic Energy Government of India (DAE), Department of Science and Technology, Government of India (DST), University Grants Commission, Government of India (UGC) and Council of Scientific and Industrial Research (CSIR), India;
Indonesian Institute of Science, Indonesia;
Istituto Nazionale di Fisica Nucleare (INFN), Italy;
Institute for Innovative Science and Technology , Nagasaki Institute of Applied Science (IIST), Japanese Ministry of Education, Culture, Sports, Science and Technology (MEXT) and Japan Society for the Promotion of Science (JSPS) KAKENHI, Japan;
Consejo Nacional de Ciencia (CONACYT) y Tecnolog\'{i}a, through Fondo de Cooperaci\'{o}n Internacional en Ciencia y Tecnolog\'{i}a (FONCICYT) and Direcci\'{o}n General de Asuntos del Personal Academico (DGAPA), Mexico;
Nederlandse Organisatie voor Wetenschappelijk Onderzoek (NWO), Netherlands;
The Research Council of Norway, Norway;
Commission on Science and Technology for Sustainable Development in the South (COMSATS), Pakistan;
Pontificia Universidad Cat\'{o}lica del Per\'{u}, Peru;
Ministry of Science and Higher Education, National Science Centre and WUT ID-UB, Poland;
Korea Institute of Science and Technology Information and National Research Foundation of Korea (NRF), Republic of Korea;
Ministry of Education and Scientific Research, Institute of Atomic Physics and Ministry of Research and Innovation and Institute of Atomic Physics, Romania;
Joint Institute for Nuclear Research (JINR), Ministry of Education and Science of the Russian Federation, National Research Centre Kurchatov Institute, Russian Science Foundation and Russian Foundation for Basic Research, Russia;
Ministry of Education, Science, Research and Sport of the Slovak Republic, Slovakia;
National Research Foundation of South Africa, South Africa;
Swedish Research Council (VR) and Knut \& Alice Wallenberg Foundation (KAW), Sweden;
European Organization for Nuclear Research, Switzerland;
Suranaree University of Technology (SUT), National Science and Technology Development Agency (NSDTA) and Office of the Higher Education Commission under NRU project of Thailand, Thailand;
Turkish Atomic Energy Agency (TAEK), Turkey;
National Academy of  Sciences of Ukraine, Ukraine;
Science and Technology Facilities Council (STFC), United Kingdom;
National Science Foundation of the United States of America (NSF) and United States Department of Energy, Office of Nuclear Physics (DOE NP), United States of America.
\end{acknowledgement}

\bibliographystyle{utphys}
\bibliography{AliHFMPbPb5dot02}

\newpage
\appendix

\section{The ALICE Collaboration}
\label{app:collab}
\begin{flushleft}

\bigskip 

S.~Acharya$^{\rm 142}$, 
D.~Adamov\'{a}$^{\rm 96}$, 
A.~Adler$^{\rm 74}$, 
J.~Adolfsson$^{\rm 81}$, 
G.~Aglieri Rinella$^{\rm 34}$, 
M.~Agnello$^{\rm 30}$, 
N.~Agrawal$^{\rm 54,9}$, 
Z.~Ahammed$^{\rm 142}$, 
S.~Ahmad$^{\rm 16}$, 
S.U.~Ahn$^{\rm 76}$
% ...

\bigskip

\bigskip 

\textbf{\Large Affiliation Notes}

\bigskip 

% ...
$^{\rm VI}$ Also at: Institute of Theoretical Physics, University of Wroclaw, Poland\\

\bigskip

\bigskip 

\textbf{\Large Collaboration Institutes}

\bigskip 

% ...
$^{9}$ Centro Fermi - Museo Storico della Fisica e Centro Studi e Ricerche ``Enrico Fermi', Rome, Italy\\
% ...
$^{16}$ Department of Physics, Aligarh Muslim University, Aligarh, India\\
% ...
$^{30}$ Dipartimento DISAT del Politecnico and Sezione INFN, Turin, Italy\\
% ...
$^{34}$ European Organization for Nuclear Research (CERN), Geneva, Switzerland\\
% ...
$^{54}$ INFN, Sezione di Bologna, Bologna, Italy\\
% ...
$^{74}$ Johann-Wolfgang-Goethe Universit\"{a}t Frankfurt Institut f\"{u}r Informatik, Fachbereich Informatik und Mathematik, Frankfurt, Germany\\
% ...
$^{76}$ Korea Institute of Science and Technology Information, Daejeon, Republic of Korea\\
% ...
$^{81}$ Lund University Department of Physics, Division of Particle Physics, Lund, Sweden\\
% ...
$^{96}$ Nuclear Physics Institute of the Czech Academy of Sciences, \v{R}e\v{z} u Prahy, Czech Republic\\
% ...
$^{142}$ Variable Energy Cyclotron Centre, Homi Bhabha National Institute, Kolkata, India\\
% ...

\bigskip 

\end{flushleft} 

\end{document}